\documentclass[12pt]{article}

\usepackage{amsmath}

\usepackage{amsfonts}

\numberwithin{equation}{section}

\def\thickone{{\rm 1\mskip-4.5mu l}}
\def\normalorder{{:}} 

\begin{document}

\title{{\bf The Orbifolds of Permutation-Type as
\\ Physical String Systems\\ at Multiples~of $\mathbf{c=26}$
  \\III.  The Spectra of $\mathbf{\hat{c}=52}$ Strings}}

\author{M.B.Halpern\footnote{halpern@physics.berkeley.edu}
\\ Department of Physics, University of California\\
 and Theoretical Physics Group,\\
  Lawrence Berkeley National Laboratory \\
University of California, Berkeley CA 94720 USA}

\maketitle

\begin{abstract}

\noindent In the second paper of this series, I obtained the twisted BRST 
systems and extended physical-state conditions of all twisted open 
and closed $\hat{c} = 52$ 
strings.
In this paper, I supplement the extended physical-state conditions 
with the explicit form of the extended (twisted) Virasoro generators of all
$\hat{c} = 52$ strings, which allows us to discuss the physical spectra of 
these systems. Surprisingly, all the $\hat{c}=52$ spectra admit an 
equivalent
description in terms of generically-unconventional Virasoro 
generators at $c=26$. This description strongly supports our prior conjecture that 
the  $\hat{c}=52$ strings are free of negative-norm states, and 
moreover shows that the spectra of some of the simpler cases are equivalent 
to those of ordinary untwisted open and closed $c=26$ strings. 

\end{abstract}

\newpage

\begin{flushleft}
{\bf \large Table of Contents}
\end{flushleft}
\begin{enumerate}
{\bf
\item[1.] Introduction

\item[2.] The Extended Virasoro Generators of $\mathbf{\hat{c}=52}$ Strings

\item[3.] First Discussion of the $\mathbf{\hat{c}=52}$ String Spectra

\item[4.] Equivalent $\mathbf{c = 26}$ Description \\ 
\hspace*{.4in} of the $\mathbf{\hat{c}=52}$ Spectra

\item[5.] Conclusions}
\end{enumerate}

\newpage

\section{Introduction}

Opening another chapter in the orbifold program [1-11,12-15], this is the third 
in a series of papers which considers the critical orbifolds of 
permutation-type as candidates for {\it new physical string systems} at higher central charge. 
In the first paper [16] of this series, we found that the twisted sectors of  
 these orbifolds are governed by new, extended (permutation-twisted) 
world-sheet gravities -- which indicate that the free-bosonic 
orbifold-string systems 
of permutation-type can be free 
of negative-norm states at  critical central charge $\hat c=26K$. 
Correspondingly-extended world-sheet 
permutation supergravities are expected in the twisted 
sectors of the superstring orbifolds of permutation-type, where 
superconformal matter lives at higher multiples of critical 
superstring central charges.

In the second 
paper [17] of the series, we found the corresponding twisted BRST systems for all sectors 
of the free-bosonic orbifolds which couple to the simple case of 
$\mathbb{Z}_{2}$-twisted permutation gravity, i.e. for all the twisted 
strings with  $\hat c=52$ matter. The new BRST systems also implied the 
following {\it extended physical-state conditions} for the physical states
$\{|\chi\rangle\}$ of each of the $\hat c=52$ strings:
\begin{subequations}
\begin{equation}\label{1.1a}
\left(\hat{L}_u \left(\left(m+\tfrac{u}{2}\right)\geq 0 \right) - 
\delta_{m+\tfrac{u}{2},0}\tfrac{17}{8}\right)|\chi \rangle  =  0, 
\hspace{2em} m\in \mathbb{Z},\quad \overline{u} = 0,1
\end{equation}
\begin{eqnarray}\label{1.1b}
\left[\hat{L}_u\left(m+\tfrac{u}{2}\right),\hat{L}_v\left(n+\tfrac{v}{2}\right) 
\right]  = \left( m-n + 
\tfrac{u-v}{2}\right)\hat{L}_{u+v}\left(m+n+\tfrac{u+v}{2}\right) \\
  \hspace{1em} +\tfrac{52}{12}\left( 
\left(m+\tfrac{u}{2}\right)\left(\left(m+\tfrac{u}{2}\right)^2-1\right)\right)\delta_{m+n+\frac{u+v}{2},0}. 
\nonumber
\end{eqnarray}
\end{subequations}
The algebra in Eq. (1.1b) is called an order-two orbifold Virasoro 
algebra (or extended, twisted Virasoro algebra) and general orbifold Virasoro algebras 
[1,18,9,12,16,17] are known to govern all the twisted sectors of the orbifolds of 
permutation-type at higher multiples of $c=26$.

The set of all $\hat c=52$ orbifold-strings is a very large class of 
fractional-moded free-bosonic string systems, 
including e.g. the twisted open-string sectors of the 
orientation orbifolds, the twisted closed-string sectors of the 
generalized $\mathbb{Z_{2}}$-permutation orbifolds and many others 
(see Refs. [16,17] and Sec. 2).  
Starting from the extended physical-state conditions (1.1) (and a 
right-mover copy of (1.1) on the same $\{|\chi\rangle\}$for the twisted closed-string sectors) this paper
begins the concrete study of the physical spectrum of each  
$\hat c=52$ string.

As the prerequisite for this analysis, I first provide in Sec. 2 the explicit 
form -- in terms of twisted matter fields -- of the extended Virasoro generators 
$\left\{\hat{L}_u\left(m+\tfrac{u}{2}\right),\overline{u}=0,1\right\}$
of all \mbox{$\hat c=52$} strings. This construction allows us to begin 
the study of the general
$\hat c=52$ string spectra in Sec. 3. The same subject is further 
considered in Sec. 4,  where I point out that all the 
$\hat c=52$ spectra admit an {\it equivalent description} in terms of 
generically-unconventional Virasoro generators at $c=26$. This 
description allows us to see clearly a number of spectral regularities which 
are only glimpsed in Sec. 3, including strong 
 further evidence that the critical orbifolds of permutation-type can be 
free of negative-norm states. Moreover, although the generic $\hat 
c=52$ spectrum is apparently new, we are able to show that some of the 
simpler spectra are equivalent to those of ordinary untwisted open 
and closed critical strings at $c=26$.

Based on these results, the discussion in Sec. 5  raises some 
interesting questions 
about these theories at the 
interacting level, and speculates on the form of the
extended physical-state conditions for more general orbifold-strings of 
permutation-type. I will return to both of these subjects in  
succeeding papers of the series.

\section{The Twisted Virasoro Generators\\
 \hspace*{.5in}  of $\mathbf{\hat{c}=52}$ Strings}
 
 As emphasized in Ref. [17], the universal form of the twisted BRST 
 systems and the extended physical-state conditions (1.1) are 
 consequences of their origin in $\mathbb{Z}_{2}$-twisted permutation 
 gravity, which governs all twisted $\hat c=52$ matter.
 
 There are however many distinct $\hat c=52$ strings, including the twisted 
 open-string sectors of the orientation orbifolds [12,13,15-17]
\begin{equation}\label{2.1}
\frac{U(1)^{26}}{H_-}, \hspace{2em} H_- = {\mathbb{Z}}_2(\mbox{w.s.}) \times H
\end{equation}
and the twisted closed-string sectors of the generalized 
$\mathbb{Z}_{2}$-permutation orbifolds [15-17]
\begin{equation}\label{2.2}
\frac{U(1)^{26}\times U(1)^{26}}{H_+}, \hspace{2em} H_+ = 
{\mathbb{Z}}_2(\mbox{perm}) \times H'
\end{equation}
as well as the generalized open-string $\mathbb{Z}_{2}$-permutation 
orbifolds and their $T$-duals [15-17]. For the orientation orbifolds in Eq. 
(2.1), I remind that $H_{-}$ is any 
automorphism group of the untwisted closed string $U(1)^{26}$ which includes 
world-sheet orientation-reversing automorphisms. Indeed the twisted 
open-string orientation-orbifold sectors correspond to the 
orientation-reversing automorphisms, which have the form
$\tau_-\times \omega, \omega \in H$  where the basic automorphism $\tau_{-}$ exchanges the 
left- and right-movers of the closed string and $\omega$ is an 
extra automorphism which acts uniformly on the left- and right-movers
of the closed string. Similarly, the automorphism group $H_{+}$ of 
the generalized $\mathbb{Z}_{2}$-permutation orbifolds in (2.2) is 
generated by elements of the form $\tau_{+}\!\times\!\omega, \omega\!\in H'$,
where the basic automorphism $\tau_{+}$ exchanges the two copies of the closed string and 
the extra automorphism $\omega$ again acts uniformly on the left- and right-movers of each 
closed string. In both cases, the extra automorphisms $\omega$ in 
$\tau\times\omega$ may or may not 
form a group (see the examples at the end of this section).

The spectra of different $\hat c=52$ strings are characterized by their extended 
(twisted) Virasoro generators,  all of which can in fact be written in the 
following {\it unified form}:
\begin{subequations}
\begin{eqnarray}\label{2.3a}
\hat{L}_u\left(m+\tfrac{u}{2}\right) = 
\tfrac{1}{4}\sum_{r,\mu\nu}\mathcal{G}^{n(r)\mu;-n(r)\nu}(\sigma) 
\sum_{v=0}^1\sum_{p\in\mathbb{Z}} \times \hspace{1.5in}\\
  \hspace{.2in}\times \normalorder\hat{J}_{n(r)\mu 
v}\left(p+\tfrac{n(r)}{\rho(\sigma)}+\tfrac{v}{2}\right)\hat{J}_{-n(r),\nu,u-v}\left(m-p-\tfrac{n(r)}{\rho(\sigma)}+\tfrac{u-v}{2}\right)\normalorder_M 
+ \nonumber \\
  \hspace{1in}+\delta_{m+\frac{u}{2},0}\hspace{.25em}\hat{\Delta}_0(\sigma) 
\hspace{3in}\nonumber
\end{eqnarray}
\begin{eqnarray}\label{2.3b}
  \left[\hat{J}_{n(r)\mu u} 
\left(m+\tfrac{n(r)}{\rho(\sigma)}+\tfrac{u}{2}\right),\hat{J}_{n(s)\nu 
v}\left(n+\tfrac{n(s)}{\rho(\sigma)}+\tfrac{v}{2}\right)\right]= 
\hspace{1.5in}\\
  \hspace{.2in}= 
2\left(m+\tfrac{n(r)}{\rho(\sigma)}+\tfrac{u}{2}\right)\delta_{n(r)+n(s),0 
\hspace{0.05em}\mbox{{\scriptsize mod}}\hspace{0.05em} \rho(\sigma)} 
\delta_{m+n+\frac{n(r)+n(s)}{\rho(\sigma)}+\frac{u+v}{2},0} 
\mathcal{G}_{n(r),\mu;-n(r),\nu}(\sigma) \nonumber
\end{eqnarray}
\begin{eqnarray}\label{2.3c}
\left[\hat{L}_u\left(m+\tfrac{u}{2}\right),\hat{J}_{n(r)\mu 
v}\left(n+\tfrac{n(r)}{\rho(\sigma)} + \tfrac{v}{2}\right) \right] 
\hspace{2.3in} \\
  = 
-\left(n+\tfrac{n(r)}{\rho(\sigma)}+\tfrac{v}{2}\right)\hat{J}_{n(r)\mu,u+v}\left(m+n+\tfrac{n(r)}{\rho(\sigma)}+\tfrac{u+v}{2}\right) 
\nonumber
\end{eqnarray}
\begin{equation}\label{2.3d}
\hat{\Delta}_0(\sigma) = \tfrac{13}{8}+\tfrac{1}{2}\sum_r \dim 
[\overline{n}(r)] 
\left(\tfrac{\overline{n}(r)}{\rho(\sigma)}-\tfrac{1}{2}\right)\left(\theta\left(\tfrac{\overline{n}(r)}{\rho(\sigma)}\geq\tfrac{1}{2}\right)-\tfrac{\overline{n}(r)}{\rho(\sigma)}\right)
\end{equation}
\begin{equation}\label{2.3e}
\sum_{r}\dim[\overline{n}(r)] = 26.
\end{equation}
\end{subequations}
Each set of extended Virasoro generators in Eq. (2.3a) satisfies the 
order-two orbifold Virasoro algebra (1.1b) at $\hat c=52$, and the 
current algebras in Eq. (2.3b) are of the type called {\it 
doubly-twisted} in the orbifold program.

For those unfamiliar with the program, I first give a short summary of the 
standard notation in the result (2.3) -- followed by the derivation of 
the result. As in the extended Virasoro generators themselves, the 
indices $u,v$ with fundamental range $\bar u,\bar v \in \{0,1\}$ describe the twist of 
the basic permutations $\tau_{\mp}$ in each $H_{\mp}$. For each extra automorphism 
$\omega(\sigma)$ in each $\tau_{\mp}\times\omega(\sigma)$, the spectral indices $\{n(r)\}$ 
and the degeneracy indices $\{\mu \equiv \mu(n(r))\}$ of each twisted 
sector $\sigma$ are determined by the so-called {\it $H$-eigenvalue 
problem} [3,5,6] 
of $\omega(\sigma)$
\begin{subequations}
\begin{equation}\label{2.4a}
{\omega(\sigma)_a}^b{U^\dag(\sigma)_b}^{n(r)\mu} = 
{U^\dag(\sigma)_a}^{n(r)\mu} e^{-2\pi i \frac{n(r)}{\rho(\sigma)}}, 
\hspace{2em} \omega(\sigma) \in H \mbox{ or } H'
\end{equation}
\begin{equation}\label{2.4b}
{\omega(\sigma)_a}^c{\omega(\sigma)_b}^dG_{cd}=G_{ab}, \hspace{2em} 
G_{ab} = G^{ab} = \left(\begin{array}{cc} -1 & 0 \\ 0 & \thickone 
\end{array}\right)
\end{equation}
\begin{equation}\label{2.4c}
a,b = 0,1,\ldots, 25,\hspace{2em} \overline{n}(r) \in (0,1,\ldots, 
\rho(\sigma)-1)
\end{equation}
\end{subequations}
where $G$ is the untwisted target-space metric of $U(1)^{26}$.
The quantity $\rho(\sigma)$ is the order of $\omega(\sigma)$ and all 
indices $\{n(r)\mu\}$ are periodic modulo $\rho(\sigma)$, with 
$\{\overline{n}(r)\}$ the 
pullback to the fundamental region and $\dim[\overline{n}(r)]$ the 
size of the subspace $\overline{n}(r)$. The index r is 
summed once over the fundamental region in Eqs. (2.3a), (2.3d) and (2.3e).
The {\it twisted metric}  $\mathcal{G}_{\mbox{\bf .}}(\sigma)$
and its inverse $\mathcal{G}^{\mbox{\bf .}}(\sigma)$ are defined in 
terms of the unitary eigenvectors $U(\sigma)$ of the H-eigenvalue 
problem 
\begin{subequations}
\begin{eqnarray}\label{2.5a}
{\mathcal G}_{n(r)\mu;n(s)\nu}(\sigma) & = & 
\chi_{n(r)\mu}\chi_{n(s)\nu}{U(\sigma)_{n(r)\mu}}^a{U(\sigma)_{n(s)\nu}}^bG_{ab} 
\\
  & = & \delta_{n(r)+n(s),0 \hspace{0.05em}\mbox{{\scriptsize 
mod}}\hspace{0.05em} \rho(\sigma)} {\mathcal G}_{n(r)\mu; 
-n(r),\nu}(\sigma)\label{2.5b}
\end{eqnarray}
\begin{eqnarray}\label{2.5c}
{\mathcal G}^{n(r)\mu; n(s)\nu}(\sigma) & = & 
\chi^{-1}_{n(r)\mu}\chi^{-1}_{n(s)\nu}G^{ab}{U^\dag(\sigma)_a}^{n(r)\mu}{U^\dag(\sigma)_b}^{n(s)\nu} 
\\
  & = & \delta_{n(r)+n(s), 0 \hspace{0.05em}\mbox{{\scriptsize 
mod}}\hspace{0.05em} \rho(\sigma)} {\mathcal G}^{n(r),\mu; -n(r),\nu}\label{2.5d}
\end{eqnarray}
\end{subequations}
where $G$ is again the untwisted metric and the $\chi$'s are 
essentially-arbitrary normalization constants. Finally, the standard 
mode normal-ordering in Eq. (2.3a) is:
\begin{eqnarray}\label{2.6}
\normalorder\hat{J}_{n(r)\mu u} 
\left(m+\tfrac{n(r)}{\rho(\sigma)}+\tfrac{u}{2}\right)\hat{J}_{n(s)\nu 
v}\left(n+\tfrac{n(s)}{\rho(\sigma)}+\tfrac{v}{2}\right)\normalorder_M 
\hspace{10em}& &  \\
 \hspace{.1in} = 
\theta\left(\left(m+\tfrac{n(r)}{\rho(\sigma)}+\tfrac{u}{2}\right)\geq 
0 \right)\hat{J}_{n(s)\nu 
v}\left(n+\tfrac{n(s)}{\rho(\sigma)}+\tfrac{v}{2}\right)\hat{J}_{n(r)\mu 
u}\left(m+\tfrac{n(r)}{\rho(\sigma)}+\tfrac{u}{2}\right) & & 
\nonumber \\
+ \theta\left(\left(m+\tfrac{n(r)}{\rho(\sigma)}+\tfrac{u}{2}\right)< 
0 \right)\hat{J}_{n(r)\mu 
u}\left(m+\tfrac{n(r)}{\rho(\sigma)}+\tfrac{u}{2}\right)\hat{J}_{n(s)\nu 
v}\left(n+\tfrac{n(s)}{\rho(\sigma)}+\tfrac{v}{2}\right). & & \nonumber
\end{eqnarray}
It follows that the quantity $\Delta_{0}(\sigma)$ in Eqs. (2.3a) and (2.3d)
\begin{subequations}
\begin{equation}\label{2.7a}
\hat{J}_{n(r)\mu 
u}\left(\left(m+\tfrac{n(r)}{\rho(\sigma)}+\tfrac{u}{2}\right)\geq 0 
\right)|0\rangle_\sigma = 0
\end{equation}
\begin{equation}\label{2.7b}
\rightarrow\quad\hat{L}_u\left(\left(m+\tfrac{u}{2}\right)\geq0\right)|0\rangle_\sigma 
= \hat{\Delta}_0(\sigma)\,\, \delta_{m+\frac{u}{2},0}|0\rangle_\sigma
\end{equation}
\end{subequations}
is the conformal weight of the scalar twist-field state $|0\rangle_\sigma$
of sector $\sigma$.

I comment briefly on the derivation of the unified form (2.3) of the 
$\hat c=52$ extended Virasoro generators. Essentially this result was given for 
the twisted open-string sectors of the non-abelian orientation 
orbifolds in Subsecs. 3.4, 3.5 of Ref. [12], and that result is easily reduced 
for our abelian case $U(1)^{26} / H_-$ in Eq. (2.1). With a right-mover copy of 
the extended Virasoro generators (and
$\overline{u} \to \overline{\hat{\jmath}} = 0,1$), the result also 
hold for the twisted closed-string sectors of the generalized 
$\mathbb{Z}_{2}$-permutation orbifolds $(U(1)^{26} \times U(1)^{26})/ H_+$ 
in Eq. (2.2).
This follows by the substitution
\begin{equation}\label{2.8}
G \hspace{0.5em}\to \hspace{0.5em}{\mathcal G}, 
\hspace{2em}\tfrac{u}{2} \hspace{0.5em}\to \hspace{0.5em} 
\tfrac{n(r)}{\rho(\sigma)}+\tfrac{u}{2}
\end{equation}
into the known results for the ordinary $\mathbb{Z}_{2}$-permutation 
orbifolds with trivial $H'$ (see Ref. [perm] and Subsec. 4.2 of Ref. 
[16]). Finally, a single copy of the unified form (2.3) holds as well for 
each twisted sector of the 
generalized open-string $\mathbb{Z}_{2}$-permutation orbifolds 
$(U(1)^{26} \times U(1)^{26})_{open}/ H_+$ and all possible 
$T$-dualizations of each of these sectors. This conclusion follows because the left-mover extended
Virasoro generators 
of the closed-string orbifolds for each $H_{+}$ are the 
input data for the construction of the correponding open-string 
orbifolds [14], 
and the twisted-current form of each set of extended Virasoro generators is 
independent of T-dualization [15]. The branes, quasi-canonical algebra 
and non-commutative geometry of the twisted open strings [13-15,16,17] depend of 
course on the particular T-dualization, but these will not be needed here. 

In what follows I will consider each twisted $\hat c= 52$ string 
separately, but the reader may find it  helpful to bear in mind the complete sector 
structure of these orbifold-string systems as labelled by the 
elements of the automorphism groups $H_{\mp}$. Given a particular extra automorphism 
$\omega_n \in H$ or $H'$ of order $n$, one may list the following low-order 
examples: 
\begin{subequations}
\begin{equation}\label{2.9a}
(1;\hspace{0.25em}\tau_{\mp})
\end{equation}
\begin{equation}\label{2.9b}
(1;\hspace{0.25em}\tau_\mp\times \omega_2)
\end{equation}
\begin{equation}\label{2.9c}
(1,\hspace{0.25em}\omega_3,\hspace{0.25em}\omega_3^2;\hspace{0.25em}\tau_\mp,\hspace{0.25em}\tau_\mp\times 
\omega_3,\hspace{0.25em}\tau_\mp\times \omega_3^2)
\end{equation}
\begin{equation}\label{2.9d}
(1,\omega_4^2;\hspace{0.25em} 
\tau_\mp\times\omega_4,\hspace{0.25em}\tau_\mp\times\omega_4^3)
\end{equation}
\begin{equation}\label{2.9e}
(1,\omega_6^2,\hspace{0.25em}\omega_6^4;\hspace{0.25em}\tau_\mp\times\omega_6,\hspace{0.25em} 
\tau_\mp\times\omega_6^3,\hspace{0.25em} \tau_\mp \times \omega_6^5).
\end{equation}
\end{subequations}
For the generalized $\mathbb{Z}_{2}$-permutation orbifolds 
$(\tau_{+})$ 
all of these sectors are twisted closed strings at $\hat c=52$, 
while all the sectors of the generalized open-string $\mathbb{Z}_{2}$-permutation 
orbifolds $(\tau_{+})$ and their T-dualizations are twisted open 
strings at $\hat c=52$. For the orientation orbifolds $(\tau_{-})$ 
the sectors before the semicolon are twisted closed strings at 
$c=26$ (which form an ordinary space-time orbifold) while the sectors after the 
semicolon are twisted open strings at $\hat c=52$. More generally, 
orientation orbifolds always contain an equal number of twisted open 
and closed strings. In all cases, the twisting is of course trivial 
for sectors corresponding to the unit element.

\section{First Discussion of \\ 
\hspace*{.4in} the $\mathbf{ \hat{c}= 52}$ String Spectra}

To frame this discussion, I remind [1] the reader that the 
Virasoro primary states of our orbifold CFT's are defined by 
the integral Virasoro subalgebra (generated by $\{\hat{L}_0(m)\}$) of 
the extended Virasoro algebra. Then the extended physical-state 
conditions (1.1a) tell us that all the physical states 
$\{|\chi\rangle\}$ of each $\hat c=52$ orbifold-string are Virasoro 
primary
\begin{equation}\label{3.1}
\hat{L}_0(m>0) |\chi\rangle = 0
\end{equation}
but only a small subset of these primary states are selected by the rest
of the physical-state conditions:
\begin{equation}\label{3.2}
\left(\hat{L}_0(0)-\tfrac{17}{8}\right)|\chi\rangle = 
\hat{L}_1\left(\left(m+\tfrac{1}{2}\right)> 0 \right)|\chi\rangle = 0.
\end{equation}
In what follows, I will refer to the $\hat{L}_0(0)$ condition in Eq. 
(3.2) as the {\it spectral condition},  since it will determine the 
allowed values of momentum-squared for each $\hat c=52$ string.

The space of physical states of each orbifold-string is then {\it much 
smaller} 
than the space of states of the underlying orbifold conformal 
field theory. For the experts, I remark in particular that the 
extended physical-state conditions generically disallow the 
characteristic sequence [19] of Virasoro primary states known as the 
principle-primary
states [1,9]. This follows first by the spectral 
condition (which fixes the conformal weight), and second because the 
physical-state condition 
$\{\hat{L}_u\left(\left(m+\frac{u}{2}\right)>0\right)\simeq 0\}$
is stronger than the principle-primary state condition [1,9]
\begin{equation}\label{3.3}
\hat{L}_u\left(m+\tfrac{u}{2}\right)|\mbox{p.p.s.}\rangle= 0, 
\hspace{2em} \overline{u} = 0,1, \hspace{1em} m>0
\end{equation}
which does not extend to $m=0$.

I turn now to concretize the spectral condition of each twisted $\hat c=52$ 
string, using the explicit form (2.3) of its extended Virasoro generators.
For this, recall [12,15] first that these generators contain in 
general two kinds of commuting zero modes (dimensionless momenta), 
namely $\{\hat{J}_{0\mu 0}(0)\}$ and 
$\{\hat{J}_{\rho(\sigma)/2,\mu,1}(0)\}$,
where the latter is relevant only when the order $\rho(\sigma)$ of $\omega(\sigma)$
is even. In what follows, I often refer to these zero modes 
collectively as $\{\hat{J}(0)\}$. It is then natural to define the 
``momentum-squared'' operator $\hat{P}^2$ as follows:
\begin{subequations}
\begin{equation}\label{3.4a}
\hat{L}_0(0) = 
\tfrac{1}{4}\left(-\hat{P}^2+\hat{R}(\sigma)\right)+\hat{\Delta}_0(\sigma)
\end{equation}
\begin{eqnarray}\label{3.4b}
\hat{P}^2 & \equiv & -\sum_{\mu,\nu} \Big{\{} {\mathcal 
G}^{0\mu;0\nu}(\sigma)\hat{J}_{0\mu0}(0)\hat{J}_{0\nu0}(0)+ \\
  & & \hspace{5em}+{\mathcal 
G}^{\frac{\rho(\sigma)}{2},\mu;-\frac{\rho(\sigma)}{2},\nu}(\sigma)
\hat{J}_{\rho(\sigma)/2,\mu,1}(0)\hat{J}_{-\rho(\sigma)/2,\nu,-1}(0) \Big{\}} \nonumber
\end{eqnarray}
\begin{eqnarray}\label{3.4c}
\hat{R}(\sigma) & \equiv & 
\left(\sum_{r,\mu,\nu}\sum_u\sum_{p\in\mathbb{Z}}\right)^\prime{\mathcal 
G}^{n(r)\mu;-n(r),\nu}(\sigma)\times \\
  & & \hspace{1em}\times\normalorder\hat{J}_{n(r)\mu 
u}\left(p+\tfrac{n(r)}{\rho(\sigma)}+\tfrac{u}{2}\right)
\hat{J}_{-n(r),\nu,-u}\left(-p-\tfrac{n(r)}{\rho(\sigma)}-\tfrac{u}{2}\right)\normalorder_M. 
\nonumber
\end{eqnarray}
\end{subequations}
Here the primed sum in the ``level-number'' operator $\hat{R}(\sigma)$ 
indicates omission of the zero modes.

With this decomposition, the spectral condition in Eq. (3.2) takes 
the simple form:
\begin{subequations}
\begin{equation}\label{3.5a}
\hat{P}^2|\chi\rangle = 
\left(\hat{P}^2_{(0)}+\hat{R}(\sigma)\right)|\chi\rangle
\end{equation}
\begin{equation}\label{3.5b}
\hat{P}^2_{(0)}\, \equiv\, 2\left(\hat{\delta}_0(\sigma)-1\right)
\end{equation}
\begin{equation}\label{3.5c}
\hat{\delta}_0(\sigma) = \sum_r \dim 
[\overline{n}(r)]\left(\tfrac{\overline{n}(r)}{\rho(\sigma)}-\tfrac{1}{2}\right)\left(\theta\left(\tfrac{\overline{n}(r)}{\rho(\sigma)}>\tfrac{1}{2}\right)-\tfrac{\overline{n}(r)}{\rho(\sigma)}\right) 
\,\geq 0.
\end{equation}
\end{subequations} 
Although I will continue the discussion primarily in this form, in 
fact  Eqs.  (3.4a) and (3.5a)
 hold only for the twisted open-string 
sectors of the orbifolds. For the twisted closed-string sectors, we  also 
have right-mover copies of 
the extended Virasoro generators (2.3), and a corresponding right-mover copy of the extended 
physical-state conditions (1.1) on the same $\{|\chi\rangle\}$.  For 
simplicity I will limit the 
discussion of these sectors here to the case of decompactified zero 
modes, for which it
is appropriate to equate the left and right movers 
\begin{equation}\label{3.6}
\hat{J}^R(0) = \hat{J}^L(0) = \tfrac{1}{\sqrt{2}}\hat{J}(0) 
\hspace{0.5em} \to\hspace{0.5em} \hat{R}^R(\sigma)=\hat{R}^L(\sigma)
\end{equation}
where the last equality is level-matching in each twisted sector.
Keeping the same definition of the operator $\hat{P}^2$ in Eq. 
(3.4b), the correct closed-string $\hat c=52$ spectral condition is then obtained 
by the substitution
\begin{equation}\label{3.7}
\hat{P}^2 \to \tfrac{1}{2}\hat{P}^2
\end{equation}
in both Eqs. (3.4a) and (3.5a). These identifications, and hence 
$\hat{P}^{2}_{(0)}\to 2\hat{P}^{2}_{(0)}$, can be used at 
any point in the discussion below to obtain the corresponding 
closed-string results.

Returning to the open-string case, one simple solution of the extended physical-state conditions is the 
ground state $|0,\hat{J}(0)\rangle_\sigma$ of twisted sector $\sigma$:
\begin{subequations}
\begin{equation}\label{3.8a}
\hat{R}(\sigma)|0,\hat{J}(0)\rangle_\sigma = 
\hat{L}_u\left(\left(m+\tfrac{u}{2}\right)>0\right)|0,\hat{J}(\sigma)\rangle_\sigma=0
\end{equation}
\begin{equation}\label{3.8b}
\hat{P}^2|0,\hat{J}(0)\rangle_\sigma = 
\hat{P}^2_{(0)}|0,\hat{J}(0)\rangle_{\sigma}, \hspace{2em} 
\hat{P}^2_{(0)}=-2+2\hat{\delta}_0(\sigma)
\end{equation}
\end{subequations}
This is the ``momentum-boosted'' twist-field state (see Eq. (2.7)) of that sector,
with ground-state mass-squared $\hat{P}^2_{(0)}$. 
Moreover Eq. (3.4c) and the commutator (2.3c) give the increments
\begin{equation}\label{3.9}
\Delta(\hat{P}^2) = \Delta(\hat{R}(\sigma))=4\left| 
m+\tfrac{\overline{n}(r)}{\rho(\sigma)}+\tfrac{u}{2}\right|
\end{equation}
obtained by adding the negatively-moded current 
\begin{equation}
\hat{J}_{n(r)\mu 
u}\left(\left(m+\tfrac{n(r)}{\rho(\sigma)}+\tfrac{u}{2}\right)< 
0\right) \nonumber
\end{equation}
to any previous state. The precise content of these excited
levels must of course be determined from the remainder of the extended
physical-state conditions.

I continue this discussion with some specific examples of $\hat c=52$ 
strings, beginning with 
the simplest twisted open-string orientation-orbifold sectors 
[12,13,15,16,17]:
\begin{subequations}
\begin{equation}\label{3.10a}
\omega= \thickone: \hspace{1em}\rho =1, \hspace{1em}\overline{n}=0, 
\hspace{1em}U = \thickone, \hspace{1em}{\mathcal G} = G, 
\hspace{1em}\hat{J}_{0au}\left(m+\tfrac{u}{2}\right)
\end{equation}
\begin{equation}\label{3.10b}
\Delta(\hat{P}^2) = 4\left|m+\tfrac{u}{2}\right|\hspace{2em} 
(\overline{u} = 0 \mbox{ is DD},\,\overline{u}=0 \mbox{ is ND})
\end{equation}
\end{subequations}
\begin{subequations}
\begin{equation}\label{3.11a}
\omega= -\thickone: \hspace{1em}\rho =2, 
\hspace{0.5em}\overline{n}=1, \hspace{0.5em}U = \thickone, 
\hspace{0.5em}{\mathcal G} = G, 
\hspace{0.5em}\hat{J}_{1au}\left(m+\tfrac{u+1}{2}\right)
\end{equation}
\begin{equation}\label{3.11b}
\Delta(\hat{P}^2) = 4\left|m+\tfrac{u+1}{2}\right|\hspace{2em} 
(\overline{u} = 0 \mbox{ is DN},\,\overline{u}=0 \mbox{ is NN}).
\end{equation}
\end{subequations}
In these cases, the extra automorphisms $\omega$ act uniformly on the 
labels $a=0,\ldots25$ and $G$ is the untwisted target space metric in Eq. (2.4b). 
Although both twisted strings have $(26+26) = 52$ matter degrees of freedom, note that each example
 has only one of the two types
of zero modes $\{\hat{J}(0)\}$: $26 DD$ zero modes 
$\{\hat{J}_{0a0}(0)\}$ for
 $\omega = \thickone$ and $26 NN$ zero modes 
 $\{\hat{J}_{\rho/2,a,1}(0)\}$ for
 $\omega = -\thickone$. In both cases, the 
 momentum-squared (3.4b) has the schematic form
\begin{equation}\label{3.12}
\hat{P}^2 = \eta^{ab}\hat{J}_a(0)\hat{J}_b(0), \hspace{2em} \eta = 
\left(\begin{array}{cc} 1 & 0 \\ 0 & - \thickone\end{array}\right)
\end{equation}
where $\eta=-G$ is the standard (west-coast) 26-dimensional target-space metric. Then we compute from 
Eqs. (3.5b) and (3.5c) that both
strings share the same tachyonic ground-state mass-squared
\begin{equation}\label{3.13}
\hat{\delta}_0(\sigma) = 0, \hspace{1em}\,\, \hat{\Delta}_0(\sigma)= 
\tfrac{13}{8}, \hspace{1em}\,\, \hat{P}^2_{(0)}= -2
\end{equation}
and the first excited state of each is massless:
\begin{equation}\label{3.14}
\hat{P}^2\left\{ 
\begin{array}{c}\hat{J}_{0a1}\left(-\tfrac{1}{2}\right) \\ 
\hat{J}_{1a0}\left(-\tfrac{1}{2}\right) \end{array}\right\} 
|0,\hat{J}(0)\rangle_\sigma = 0 \hspace{0.5em}\mbox{ for 
}\hspace{0.5em} \omega = \left\{\begin{array}{c} \thickone \\ 
-\thickone.\end{array}\right.
\end{equation}
For this level, I have checked that the $\hat{L}_1\left(\tfrac{1}{2}\right) \simeq 0$
gauge eliminates the longitudinal parts of the $26$-dimensional 
``photons'', and moreover the $\hat{L}_1\left(\tfrac{1}{2}\right)$ and $\hat{L}_0\left(1\right)$
gauges together eliminate the negative-norm states at the next level:
\begin{equation}\label{3.15}
\left(\alpha\left(\hat{J}\left(-\tfrac{1}{2}\right)\right)^2+\beta\hat{J}(-1)\right)|0,\hat{J}(0)\rangle_\sigma, 
\hspace{2em} \hat{P}^2 = 2.
\end{equation}
Since the increments $\Delta(P^{2})$ in Eqs. (3.10b) and (3.11b) are 
even integers, we are led to suspect that the spectra of these two 
twisted $\hat c=52$ strings are nothing but the spectrum of an ordinary
open $c=26$ string in disguise \footnote{The spectra of these two $\hat 
c=52$ strings look even more familiar in terms of the dimensionful 
momenta $k \equiv\hat{J}(0)/\sqrt{2\alpha'_0}$, where $\alpha'_{0}$ is 
the conventional open-string Regge slope.}. I will return to this question in the 
following section.

A larger subset of twisted $\hat c=52$ strings is the following. For a 
particular twisted sector $\sigma$, suppose that 
$\omega=\pm\thickone$ acts uniformly on a set of $d$ labels 
$a=0,1,\ldots d-1, d\geq 4$ while a non-trivial element 
$\omega(\mbox{perm})$ of some permutation group acts non-trivially on the 
other $26-d$ spatial labels. Then Eqs. (2.4),(2.5) and standard 
results [3,5-7,9] in the orbifold program  give the following
 explicit form of the  extended Virasoro
generators (2.3) in this sector:
\begin{subequations}
\begin{eqnarray}\label{3.16a}
& &\hat{L}_u\left(m+\tfrac{u}{2}\right) = \delta_{m+\tfrac{u}{2},0} 
\hspace{0.25em}\Delta_0(\sigma)+ \\
  & &\hspace{1em}+\tfrac{1}{4}G^{ab}_{(d)}\sum_v \sum_p \normalorder 
\hat{J}_{\epsilon a 
v}\left(p+\tfrac{v+\epsilon}{2}\right)\hat{J}_{-\epsilon,b,u-v}\left(m-p+\tfrac{u-v-\epsilon}{2}\right)\normalorder_M 
+\nonumber \\
& & \hspace{1em}+\tfrac{1}{4}\sum_v\sum_j\tfrac{1}{f_j(\sigma)} 
\sum_{\hat{\jmath}=0}^{f_j(\sigma)-1}\sum_p \times \nonumber \\
  & & \hspace{5em}\times\normalorder 
\hat{J}_{\hat{\jmath}jv}\left(p+\tfrac{\hat{\jmath}}{f_j(\sigma)}+\tfrac{v}{2}\right)\hat{J}_{-\hat{\jmath},j,u-v}\left(m-p-\tfrac{\hat{\jmath}}{f_j(\sigma)}+\tfrac{u-v}{2}\right)\normalorder_M 
\nonumber
\end{eqnarray}
\begin{equation}\label{3.16b}
\hat{\delta}_0(\sigma) = \sum_j 
\sum_{\hat{\jmath}=0}^{f_j(\sigma)-1}\left(\tfrac{\hat{\jmath}}{f_j(\sigma)}-\tfrac{1}{2}\right)\left(\theta\left(\tfrac{\hat{\jmath}}{f_j(\sigma)}> 
\tfrac{1}{2}\right)-\tfrac{\hat{\jmath}}{f_j(\sigma)}\right)\geq 0
\end{equation}
\begin{equation}\label{3.16c}
\sum_j f_j(\sigma) = 26 - d, \hspace{2em} 4\leq d \leq 26.
\end{equation}
\end{subequations}
Here $\epsilon = 0$ or 1 for $\omega = \thickone$ or $-\thickone$, 
$G^{(d)}$ is the restriction of the flat target-space metric (2.4b) to 
the first $d$ labels,
$f_j(\sigma)$ is the size of the $j$th cycle in 
$\omega(\mbox{perm})$, and the previous cases with 
$\hat{\delta}_0(\sigma)=0$ are included when $d=26$. The half-integer 
moded currents in 
the second term of (3.16) satisfy the twisted current algebra  (2.3b) with 
 $\mathcal G \to G^{(d)}$. For the 
permutation-twisted currents in the last term of (3.16), I have used  the standard relation 
$(n(r)/\rho(\sigma))=(\hat{\jmath}/f_{j}(\sigma))$ and (the inverse of) the
twisted metric [3,5-7,9]
\begin{equation}\label{3.17}
{\mathcal G}_{\hat{\jmath}j;\hat{l}l}(\sigma) = 
\delta_{jl}f_j(\sigma) \delta_{\hat{\jmath}+\hat{l},\hspace{0.05em} 0 
\hspace{0.05em}\mbox{{\scriptsize mod}}\hspace{0.05em} f_j(\sigma)}
\end{equation}
which also determines the twisted current algebra (2.3b) for these 
currents. Using Eq. 
(3.16b), we see that the non-trivial element of $\mathbb{Z}_{2}$ on 
two labels also gives $\hat{\delta}_0(\sigma) = 0$ and a $\hat{P}^2_{(0)} = -2$
ground state, but a non-trivial element of $\mathbb{Z}_{3}$ on three 
labels gives a slightly-raised ground state
\begin{equation}\label{3.18}
\hat{\delta}_0(\sigma) = \tfrac{1}{9}, \hspace{1em}\,\, 
\hat{\Delta}_0(\sigma) = \tfrac{121}{72}, \hspace{1em}\,\, 
\hat{P}^2_{(0)} = -\tfrac{16}{9}
\end{equation}
and no photons. 

Given the cycle-structure $\{f_j(\sigma)\}$ of
any extra automorphism $w(\mbox{perm})$ (see e.g. Eq. (3.4) 
of Ref. [16]), it is straightforward to evaluate the sum in Eq. (3.16b).  As an  
illustration, one finds the simple tachyonic ground-state mass-squares
\begin{equation}\label{3.19}
5\leq (d = \mbox{prime}) \leq 23:\quad \hat{P}^{2}_{(0)} = -\frac{1}{12}(d-2 + 
\frac{1}{26-d}) 
\end{equation}
in twisted sectors which correspond to the action of any non-trivial element of the cyclic group 
$\mathbb{Z}_{\lambda}$ of prime order on $3\leq (\lambda = 
26-d)\leq 21$ spatial 
labels. The result (3.19) includes Eq. (3.18) when $d=23$, but does 
not extend to the cases $d=26,24$ with $\hat{P}_{(0)}^{2}=-2$ discussed above. I remind that this 
result applies only to the open orbifold-strings, while twice these values of $\hat{P}^{2}_{(0)}$ are 
obtained for the closed-string versions. 

 Further analysis of the $\hat c=52$ strings, 
including the ``larger subset'' of examples (3.16), is found in the 
following section.

\section{Equivalent $\mathbf{c = 26}$ Description \\
 \hspace*{.5in} of the $\mathbf{\hat{c} = 52}$ Spectra}
 
In fact, there exists an entirely equivalent description of all the 
$\hat c=52$ string spectra in terms of generically-unconventional Virasoro generators at $c=26$.
 
To obtain the $c=26$ description, I first define the relabelled 
(unhatted) operators 
\begin{subequations}
\begin{equation}\label{4.1a}
J_{n(r)\mu}\left(2m+u+\tfrac{2n(r)}{\rho(\sigma)}\right)\hspace{0.5em}\equiv 
\hspace{0.5em}\hat{J}_{n(r)\mu 
u}\left(m+\tfrac{n(r)}{\rho(\sigma)}+\tfrac{u}{2}\right), 
\hspace{1em} \overline{u} = 0,1
\end{equation}
\begin{equation}\label{4.1b}
L(2m+u) \hspace{0.5em}\equiv 
\hspace{0.5em}2\hat{L}_u\left(m+\tfrac{u}{2}\right)-\tfrac{13}{4}\delta_{m+\tfrac{u}{2},0}
\end{equation}
\end{subequations}
in terms of the hatted operators above. This $1\!\!-\!\!1$ map is recognized 
as a modest generalization of (the inverse of) the order-two 
orbifold-induction procedure of Borisov, Halpern and Schweigert [1]. Since
$M\equiv 2m+u, \overline{u}=0,1$ covers the integers once, we then 
find from (2.3) the explicit form of the $c=26$ generators:
\begin{subequations}
\begin{eqnarray}\label{4.2a}
&L&\!\!\!\!\!(M) = \,\,\,\hat{\delta}_0(\sigma)\delta_{M,0}+ \hspace{3.5in}\\
  &+&\!\!\!\!\!\tfrac{1}{2}\sum_{r,\mu,\nu}{\mathcal G}^{n(r)\mu; 
-n(r),\nu}(\sigma)\sum_{Q \in \mathbb{Z}}\normalorder 
J_{n(r)\mu}\left(Q+\tfrac{2n(r)}{\rho(\sigma)}\right) 
J_{-n(r),\nu}\left(M-Q-\tfrac{2n(r)}{\rho(\sigma)}\right) 
\normalorder_M\nonumber
\end{eqnarray}
\begin{equation}\label{4.2b}
\hat{\delta}_0(\sigma) = \sum_r 
\dim[\overline{n}(r)]\left(\tfrac{\overline{n}(r)}{\rho(\sigma)}-\tfrac{1}{2}\right)\left(\theta\left(\tfrac{\overline{n}(r)}{\rho(\sigma)}> 
\tfrac{1}{2}\right)-\tfrac{\overline{n}(r)}{\rho(\sigma)}\right)
\end{equation}
\begin{equation}\label{4.2c}
[L(M),L(N)] = (M-N)L(M+N)\,+\tfrac{26}{12}M(M^2-1)\delta_{M+N,0}
\end{equation}
\begin{equation}\label{4.2d}
\left[L(M),J_{n(r)\mu}\left(N+\tfrac{2n(r)}{\rho(\sigma)}\right)\right] 
= 
-\left(N+\tfrac{2n(r)}{\rho(\sigma)}\right)J_{n(r)\mu}\left(M+N+\tfrac{2n(r)}{\rho(\sigma)}\right)
\end{equation}

\begin{eqnarray}\label{4.2e}
\left[J_{n(r)\mu}\left(M+\tfrac{2n(r)}{\rho(\sigma)}\right),J_{n(s)\nu}\left(N+\tfrac{2n(s)}{\rho(\sigma)}\right)\right] 
\hspace{2.3in}\\
  \!\!= \delta_{n(r)+n(s),0 \hspace{0.05em}\mbox{{\scriptsize 
mod}}\hspace{0.05em} 
\rho(\sigma)}\delta_{M+N+2\left(\frac{n(r)+n(s)}{\rho(\sigma)}\right),0}{\mathcal 
G}_{n(r)\mu;-n(r),\nu}(\sigma).\nonumber
\end{eqnarray}
\end{subequations}
The expression (4.2b) for $\hat{\delta}_{0}(\sigma)$ is the same as 
above, and the mode-normal ordering in Eq. (4.2a)
\begin{eqnarray}\label{4.3}
\normalorder 
J_{n(r)\mu}\left(M+\tfrac{2n(r)}{\rho(\sigma)}\right)J_{n(s)\nu}\left(N+\tfrac{2n(s)}{\rho(\sigma)}\right)\normalorder_M 
\hspace{2in}\\
  = \theta\left(\left(M+\tfrac{2n(r)}{\rho(\sigma)}\right)\geq 0 
\right) J_{n(s)\nu}\left(N+\tfrac{2n(s)}{\rho(\sigma)}\right) 
J_{n(r)\mu}\left(M+\tfrac{2n(r)}{\rho(\sigma)}\right) \nonumber\\
+\theta \left(\left(M+\tfrac{2n(r)}{\rho(\sigma)}\right)< 0 \right) 
J_{n(r)\mu}\left(M+\tfrac{2n(r)}{\rho(\sigma)}\right) 
J_{n(s)\nu}\left(N+\tfrac{2n(s)}{\rho(\sigma)}\right)\nonumber
\end{eqnarray}
follows from the $\hat c=52$ ordering (2.6) because the map (4.1) preserves 
the sign of all arguments.

I emphasize that the $c=26$ Virasoro generators in Eq. (4.2) are
{\it generically-unconventional} because the twisted matter is now summed over the 
 fractions $\{2n/\rho\}$ instead of the conventional orbifold-fractions 
$\{n/\rho\}$. This distortion of the ``extra twist'' is the price we must 
pay in order to unwind the ``basic twist'' associated to the 
basic permutations $\tau_{\mp}$ of $H_{\mp}$.

The map (4.1) also tells us that the $\hat c=52$ momenta $\{{\hat 
J}(0)\}$ and the $c=26$ momenta $\{J(0)\}$ are identical, and we may 
record
\begin{subequations}
\begin{equation}\label{4.4a}
J(0) = \hat{J}(0): \hspace{1em} J_{0\mu}(0) = \hat{J}_{0\mu0}(0), 
\hspace{1em} J_{\rho(\sigma)/2, \mu}(0) = 
\hat{J}_{\rho(\sigma)/2,\mu,1}(0)
\end{equation}
\begin{eqnarray}\label{4.4b}
P^2 & = & \hat{P}^2 \\
  & = &  -\sum_{\mu,\nu} \Big{\{} {\mathcal G}^{0\mu: 0 \nu}(\sigma) 
J_{0\mu}(0) J_{0\nu}(0)+  \nonumber\\
  & & \hspace{5em}+ {\mathcal G}^{\frac{\rho(\sigma)}{2},\mu; 
-\frac{\rho(\sigma)}{2},\nu}(\sigma) 
J_{\rho(\sigma)/2,\mu}(0)J_{-\rho(\sigma)/2,\nu}(0) \Big{\}} \nonumber
\end{eqnarray}
\end{subequations}
where the $\hat c=52$ form of ${\hat P}^{2}$ was given in Eq. (3.4b).
Similarly, the  ``level-number'' operator $R(\sigma)$ in the 
decomposition of $L(0)$ is the same
\begin{subequations}
\begin{equation}\label{4.5a}
L(0) = -\tfrac{1}{2}\left(P^2 +R(\sigma)\right)+\hat{\delta}_0(\sigma)
\end{equation}
\begin{eqnarray}\label{4.5b}
R(\sigma) & = & \hat{R}(\sigma) \\
  & = & \left(\sum_{r,\mu,\nu}\sum_{Q\in\mathbb{Z}}\right)' {\mathcal 
G}^{n(r)\mu; - n(r),\nu}(\sigma) \times \nonumber \\
  & & \hspace{3em} \times\normalorder 
J_{n(r)\mu}\left(Q+\tfrac{2n(r)}{\rho(\sigma)}\right)J_{-n(r),\nu}\left(-Q-\tfrac{2n(r)}{\rho(\sigma)}\right)\normalorder_M 
\nonumber
\end{eqnarray}
\end{subequations}
where the $\hat c=52$ form of $\hat{R}(\sigma)$ was given in Eq. (3.4c).

By itself, the inverse orbifold-induction procedure (4.1) is only a relabelling of 
the operators of the permutation-orbifold CFT's. The central point of 
this discussion however is that {\it for the orbifold-string theories} 
-- restricted by the extended physical state conditions (1.1) -- the map also gives us 
a completely {\it equivalent} $c=26$ description of the  
physical spectrum of each $\hat c=52$ 
orbifold-string. Indeed, it is easily checked that both components $\bar u=0,1$
 of the $\hat c=52$ 
extended physical-state condition (1.1a) map directly onto the simpler 
and in fact conventional physical-state condition
\begin{equation}\label{4.6}
L(M\geq 0 )|\chi\rangle = \delta_{M,0}|\chi\rangle
\end{equation}
in the $26$-dimensional description! A right- mover copy of Eq. (4.6) 
on the same physical states $\{|\chi\rangle\}$ is similarly obtained 
in the equivalent $c=26$ description of the closed orbifold-strings.

I emphasize that the physical 
states $\{|\chi\rangle\}$ of the $26$-dimensional description (4.6) are
 exactly the original physical states (1.1a) of 
the $\hat c=52$ string.  Indeed, each physical state $|\chi\rangle$ can be regarded as invariant 
under the map, or each can now be rewritten in 
$26$-dimensional form. In further detail, Eqs. (4.5) and (4.6) give the same spectral 
condition $P^2 \simeq P^2_0 +R(\sigma)$, the same physical ground 
state \footnote{Although it is not directly relevant in either 
description of the $\hat c=52$ strings, one notes that the conformal 
weight of the scalar twist-field state $|0\rangle_\sigma$ of
sector $\sigma$ has now 
shifted from $\hat{\Delta}_0(\sigma)$ to $\hat{\delta}_0(\sigma)$
in the $c=26$ description.}
\begin{equation}\label{4.7}
|0,J(0)\rangle_\sigma \equiv |0,\hat{J}(0)\rangle_\sigma, 
\hspace{1em} P^2_0 = \hat{P}^2_0 = -2+2\hat{\delta}_0(\sigma)
\end{equation}
and each negatively-moded hatted current in any physical state can
be replaced according to Eq. (4.1a) by the corresponding unhatted current mode.
 Note finally that the commutator
 (4.2d) and the decomposition (4.5a) give the $26$-dimensional increment
\begin{equation}\label{4.8}
\Delta(\hat{P}^2) = \Delta(R(\sigma)) = 2 
\left|M+\tfrac{2n(r)}{\rho(\sigma)}\right|
\end{equation}
which results from the addition of $J_{n(r)\mu}\left( \left(M+\tfrac{2n(r)}{\rho(\sigma)}\right)<0\right)$
to any previous state. With $M=2m+n$, these are recognized as the 
same increments (3.9) obtained in the $\hat c=52$ description.

As simple examples, consider the ``larger subset'' (3.16) of $\hat c=52$ 
strings -- whose equivalent $c=26$ physical state condition (4.6) now involves the 
following subset of the $c=26$ Virasoro generators (4.2):
\begin{subequations}
\begin{eqnarray}\label{4.9a}
L(M)  =  \delta_{M,0}\hat{\delta}_0(\sigma) 
+\tfrac{1}{2}G^{ab}_{(d)}\sum_{Q\in \mathbb{Z}}\normalorder 
J_{\epsilon a}(Q+\epsilon)J_{-\epsilon,b} 
(M-Q-\epsilon)\normalorder_M +\hspace{1em}\nonumber\\
   +\tfrac{1}{2}\sum_j \frac{1}{f_j(\sigma)} 
\sum_{\hat{\jmath}=0}^{f_j(\sigma)-1} \sum_{Q\in 
\mathbb{Z}}\normalorder 
J_{\hat{\jmath}j}\left(Q+\tfrac{2\hat{\jmath}}{f_j(\sigma)}\right)J_{-\hat{\jmath},j}\left(M-Q-\tfrac{2\hat{\jmath}}{f_j(\sigma)}\right)\normalorder_M 
\end{eqnarray}
\begin{equation}\label{4.9b}
\hat{\delta}_0(\sigma) = \tfrac{1}{4}\sum_j 
\sum_{\hat{\jmath}=0}^{f_j(\sigma)-1}\left(\tfrac{2\hat{\jmath}}{f_j(\sigma)}-1\right)\left(2\theta\left(\tfrac{2\hat{\jmath}}{f_j(\sigma)}>1\right)-\tfrac{2\hat{\jmath}}{f_j(\sigma)}\right)
\end{equation}
\begin{equation}\label{4.9c}
a,b=0, \ldots, d-1, \hspace{1em} \sum_j f_j(\sigma) = 26 - d, 
\hspace{1em} 4\leq d \leq 26.
\end{equation}
\end{subequations}
Recall for the larger subset that $\epsilon=0,1$ corresponds in the 
symmetric theory to the action of the extra automorphism
$\omega = \pm \thickone$ on the first $d\geq4$ labels $\{a\}$, while
$f_j(\sigma)$ is the length of the $j$-th cycle of the extra 
permutation $\omega(\mbox{perm})$
which acts on the remaining $26-d$ spatial labels. Shifting the 
dummy integer $Q$ by the integer $\epsilon$, we note that the second term in
Eq. (4.9a) is a set of {\it ordinary} Virasoro generators for $d$ 
 untwisted 
bosons with the ordinary current algebra
\begin{equation}\label{4.10}
 \quad [J_{ a}(Q),J_{b}(P)] = 
G^{(d)}_{ab}Q\delta_{Q+P,0}
\end{equation}
for both values of $\epsilon$. The currents in the third term satisfy the twisted current 
algebra (4.2e) with the permutation-twisted metric (3.17), and the value of $\hat 
{\delta}_{0}(\sigma)$ in Eq. (4.9b) is only a slightly-rewritten form of that given 
in Eq. (3.16b).

 We are now in a position to confirm our suspicions in the previous 
section about the 
simplest orbifold-strings, described earlier at $\hat c=52$ by the extended Virasoro 
generators:
\begin{eqnarray}\label{4.11}
\hat{L}_u\left(m+\tfrac{u}{2}\right) & = & 
\tfrac{1}{4}G^{ab}\sum_v\normalorder\hat{J}_{\epsilon a 
v}\left(p+\tfrac{v+\epsilon}{2}\right)\hat{J}_{-\epsilon, b, 
u-v}\left(m-p+\tfrac{u-v-\epsilon}{2}\right)\normalorder_M \nonumber
\\
  & & \hspace{1em}+\tfrac{13}{8}\delta_{m+\frac{u}{2},0}, \hspace{2em} 
\overline{u} = 0,1,\,\,\, \epsilon =0,1.
\end{eqnarray}
These are now equivalently described by the choice $d=26$ in Eq. (4.9),
in which case only the second (ordinary) term of Eq. (4.9a) is non-zero --
and then the equivalent physical-state condition (4.6) verifies that 
the physical spectrum of each of these particular twisted $\hat c=52$ strings
is indeed equivalent to that of an ordinary untwisted $c=26$ string!  
These cases include the open-string 
orientation-orbifold sectors corresponding to 
${\tau}_{-}\times(\omega=\pm\thickone)$ 
in Eq. (3.10) and their T-duals, as well as the  twisted closed-string
sectors of the generalized $\mathbb{Z}_{2}$-permutation orbifolds corresponding 
to ${\tau}_{+}\times(\omega=\pm\thickone)$.

Additionally, consider the following special cases of the extended Virasoro 
generators (3.16)  at $\hat c=52$ 
\begin{eqnarray}\label{4.12}
\hat{L}_u\left(m+\tfrac{u}{2}\right) = 
\delta_{m+\frac{u}{2},0}\tfrac{13}{8}+\hspace{20em} \\
   +\tfrac{1}{4}G^{ab}_{(24)}\sum_v\sum_p\normalorder\hat{J}_{\epsilon 
a v}\left(p+\tfrac{v+\epsilon}{2}\right)\hat{J}_{-\epsilon,b,u-v}\left(m-p+\tfrac{u-v-\epsilon}{2}\right)\normalorder_M 
+\nonumber \\
  +\tfrac{1}{8}\sum_v\sum_{\hat{\jmath}=0}^1 
\sum_p\normalorder\hat{J}_{\hat{\jmath}v}\left(p+\tfrac{\hat{\jmath}+v}{2}\right)\hat{J}_{-\hat{\jmath},u-v}\left(m-p+\tfrac{u-v-\hat{\jmath}}{2}\right)\normalorder_M 
\nonumber
\end{eqnarray}
which result when the extra automorphism in the symmetric theory acts as $\omega=\pm\thickone$ 
on the first $d=24$ labels and
 the non-trivial element of a $\mathbb{Z}_{2}$ on the remaining $2$ spatial 
 labels.  I have noted in Sec. 3 that $\hat{\delta}_{0}(\sigma)=0$ for 
 these cases as well, and indeed the equivalent 
 $c=26$
 description (4.6) and (4.9) at $d=24$ now shows 
 that the open and closed  
orbifold- strings of this type also have the spectrum of ordinary 
untwisted $c=26$ strings. The common thread for the orbifold-strings in 
Eqs. (4.10) and (4.12) is that 
they are at most half-integer moded, so that the shift $\{n/\rho\} \to 
\{2n/\rho\}$
gives integer moding in the $c=26$ description.

 Beyond these simple cases, the $\hat c=52$ strings are apparently 
 new -- with $\hat{\delta}_0(\sigma) \neq 0$, unfamiliar ground-state 
 mass-squares, and fractional 
 modeing (and increments) in either description.

\section{Conclusions}

We have discussed the physical spectrum of the general  $\hat c=52$ orbifold-string, as 
well as an equivalent but unconventionally-twisted $c=26$ description of 
the twisted $\hat c=52$ matter. The 
equivalent $c=26$ description holds {\it only} for the orbifold-string 
theories -- restricted by the extended physical-state conditions 
(1.1) -- 
and {\it not} in the larger Hilbert space of the underlying orbifold conformal 
field theories.

In general we have found that the spectra of these orbifold-string 
systems are unfamiliar. One 
simple and unexpected conclusion however is that, as string theories restricted by 
the extended physical-state conditions, the single 
twisted $\hat c=52$ sector of each of the simplest orbifolds of permutation-type
(see Eq. (2.9))
\begin{subequations}
\begin{equation}\label{5.1a}
(1;\hspace{0.25em}\tau_\mp)
\end{equation}
\begin{equation}\label{5.1b}
(1;\hspace{0.25em}\tau_\mp\times\omega_2),\hspace{2em} \omega_2^2 = 1
\end{equation}
\end{subequations}
have the {\it same} physical spectra as ordinary untwisted $c=26$ strings. 
No such equivalence is found of course in the half-integer moded Hilbert space
of the full orbifold CFT's.
The list in Eq. (5.1) includes the simplest orientation 
orbifolds (with $\tau_{-}$) and their T-duals, as well as the simplest generalized 
$\mathbb{Z}_{2}$-permutation orbifolds 
(with $\tau_{+}$).

For the simplest orientation orbifolds in particular, the 
string theories in Eq. (5.1) consist of an ordinary unoriented closed string 
(the unit element) 
at $c=26$ 
and a $\hat c=52$ twisted open string whose physical spectrum is 
equivalent to that of an ordinary untwisted $c=26$ critical open string. 
Since both the closed- and open-string spectra of these simple 
orientation orbifolds are equivalent to those of the archtypal 
orientifold (without Chan-Paton factors), 
we are led to suspect that {\it orientation orbifolds include orientifolds}.
I will return in the next paper of this series to consider this question 
at the {\it interacting level}, where we will also be able to ask about 
the decoupling of null physical states.
Following that, I will consider in a succeeding paper  the corresponding 
situation and modular invariance for the simplest permutation 
orbifold-string systems.

More generally, we have seen that there are many other orientation 
orbifolds, open-string $\mathbb{Z}_{2}$-permutation orbifolds and generalized 
$\mathbb{Z}_{2}$-permutation orbifolds whose $\hat c=52$ spectra show fractional 
modeing in both the $\hat c=52$ and the $c=26$ descriptions. These 
include in particular the orbifolds in Eq.  (2.9) when the order n of 
the extra automorphism is greater than two.

There is more to say about no-ghost theorems for the general twisted 
$\hat c=52$ string. The original intuition [16] was that the doubled 
gauges $\bar{u}=0,1$ of the extended physical state condition (1.1) 
could remove the doubled set of negative-norm states (time-like modes)
of the $\hat c=52$ strings -- which are also associated with $\bar 
{u}=0,1$. For the simplest $\hat c=52$ strings in Eq. (5.1), this 
intuition is certainly born out [20]. More generally, the equivalent $c=26$
description of each spectrum shows that both aspects of the doubling 
are indeed eliminated at the same time, leaving us with the 
conventional physical state condition (4.6) and only a single set of 
time-like modes. This is clearly visible in the set of examples (4.9), 
where the only time-like modes ($a=0$) are included in the second term.
For the general $\hat c=52$ string, the reader should bear in mind that
the twisted metric $\mathcal{G}$ in Eq. (4.2) is only a unitary 
transformation (2.5) of the untwisted metric $G$ with a single time-like 
direction. Although not yet a proof, and illustrated here only for 
$\hat c=52$, I consider this a 
stronger form of the original arguments [16] that all the critical orbifolds of 
permutation-type should  be free of negative-norm states.

The next question I wish to address is the following: I have emphasized
that the equivalent $c=26$ Virasoro generators (4.2) are 
generically-unconventional, being summed over the matter-field fractions 
$\{2n/\rho\}$
instead of the conventional orbifold fractions $\{n/\rho\}$, but are they 
actually new Virasoro generators?  I do not know the answer to
this question in general, but at least some of them can in fact be 
re-expressed by further mode-relabeling in terms of more familiar Virasoro generators.
As examples, consider the special case of the ``larger subset'' (4.9) 
when $\omega(\mbox{perm})$ is one of the elements of order $\lambda$ 
of each cyclic group $\mathbb{Z}_{\lambda}$. (These are the particular,
single-cycle elements 
of $\mathbb{Z}_{\lambda}$ with  $f_0(\sigma) = \lambda$.) When 
$\lambda$ is odd, one finds that the first and third terms of (4.9) 
can in fact be re-expressed in terms of the conventional Virasoro 
generators associated to a twisted sector of an ordinary cyclic 
permutation orbifold $U(1)^\lambda/\mathbb{Z}_\lambda$ [christ]
\begin{eqnarray}\label{5.2}
L_\lambda(M) & = & 
\tfrac{1}{2\lambda}\sum_{\hat{\jmath}=0}^{\lambda-1}\sum_{Q\in\mathbb{Z}}\normalorder 
J_{\hat{\jmath}}\left(Q+\tfrac{\hat{\jmath}}{\lambda}\right)J_{-\hat{\jmath}}\left(M-Q-\tfrac{\hat{\jmath}}{\lambda}\right)\normalorder_M 
\\
  & & 
\hspace{1em}+\delta_{M,0}\tfrac{1}{24}\left(\lambda-\frac{1}{\lambda}\right), 
\hspace{2em} c = \lambda = 2 l+1 \nonumber
\end{eqnarray}
where I have relabeled the currents $J_{\hat{\jmath}}\equiv J_{\hat{\jmath}0}$.
To obtain this result from (4.9), one needs the fact that $\{2\hat{\jmath}/\lambda\} \simeq \{\hat{\jmath}/\lambda\}$
modulo the integers when $\lambda$ is odd. This observation is 
consistent with the ground-state mass-squares for prime $\lambda$ in Eq. 
(3.19). When $\lambda$ is even, I 
have also checked that the first and third terms of (4.9) can be 
re-expressed as the sum of two identical commuting Virasoro generators 
of this type
\begin{equation}\label{5.3}
L_\lambda(M) = 
L_{\frac{\lambda}{2}}(M)+\tilde{L}_{\frac{\lambda}{2}}(M), 
\hspace{2em}c = \lambda = 2l
\end{equation}
each of which is associated to a twisted sector of $U(1)^{\lambda/2}/\mathbb{Z}_{\lambda/2}$.
This result is also obtained by relabeling the modes modulo the 
integers, and provides us with another way to understand that the 
ground-state  mass-squared is unshifted when $\omega(\mbox{perm})$ is 
the non-trivial element of a $\mathbb{Z}_{2}$. 

My final remark is a conjecture, that the extended physical-state 
conditions for the twisted strings at $\hat c=26\lambda, \lambda \,\mbox{prime}$ will in fact read 
\begin{subequations}
\begin{equation}\label{5.4a}
\left(\hat{L}_{\hat{\jmath}}\left(\left(m+\tfrac{\hat{\jmath}}{\lambda}\right)\geq 
0 
\right)-\delta_{m+\frac{\hat{\jmath}}{\lambda},0}\hat{a}_\lambda\right)|\chi\rangle 
= 0, \hspace{1em} \hat{\jmath} = 0,1, \ldots, \lambda -1
\end{equation}
\begin{equation}\label{5.4b}
\hat{a}_\lambda \equiv \frac{13\lambda^2-1}{12\lambda}
\end{equation}
\begin{eqnarray}\label{5.4c}
\left[\hat{L}_{\hat{\jmath}}\left(m+\tfrac{\hat{\jmath}}{\lambda}\right),\hat{L}_{\hat{l}}\left(n+\tfrac{\hat{l}}{\lambda}\right)\right] 
\!\!& = &\!\! 
\left(m-n+\tfrac{\hat{\jmath}-\hat{l}}{\lambda}\right)\hat{L}_{\hat{\jmath}+\hat{l}}
\left(m+n+\tfrac{\hat{\jmath}+\hat{l}}{\lambda}\right)+
\\
  & & 
+\,\,\tfrac{26\lambda}{12}\left(m+\tfrac{\hat{\jmath}}{\lambda}\right)\left(\left(m+\tfrac{\hat{\jmath}}{\lambda}\right)^2-1\right)\delta_{m+n+\frac{\hat{\jmath}+\hat{l}}{\lambda},0} 
\nonumber
\end{eqnarray}
\end{subequations}
where Eq. (5.4c) is an orbifold Virasoro algebra [1,18,9] of order $\lambda$. 
This form includes the correct generators $\{\hat{L}_{\hat{\jmath}}\}$ corresponding to the 
classical extended Polyakov constraints of Ref. [16], and 
includes the correct value $\hat{a}_2 =17/8$ studied here for the $\hat c=52$ 
strings. I obtained the system (5.4) by requiring (as we now know for 
$\lambda=2$) that it map by the inverse of the order-$\lambda$ 
orbifold-induction procedure [1] 
to the conventional physical-state condition (4.6) with $\hat{a}_1 = 1$ 
at $c=26$. One way to test this conjecture would be the construction 
of the corresponding twisted BRST systems [17] for these higher values of 
$\hat c$.

Extensions to include winding number and twisted $B$ fields at $\hat 
c=52$ are also deferred to another time and place.

\section*{Acknowledgements}

For helpful information,  discussions and encouragement, I thank L. 
Alvarez-Gaum$\acute{e}$, K. Bardakci, I. Brunner,
J. de Boer, D. Fairlie, O. Ganor, E. Gimon, C. Helfgott, E. Kiritsis, R. 
Littlejohn, S. Mandelstam, J. McGreevy, N. Obers, A. Petkou, E. 
Rabinovici, V. Schomerus, K. Schoutens, C. Schweigert and E. Witten. This work was 
supported in part by the Director, Office of Energy Research, Office of 
High Energy and Nuclear Physics, Division of High Energy Physics of the U.S.
Department of Energy under Contract DE-AC02-O5CH11231 and in part by the National
Science Foundation under grant PHY00-98840.

\end{document}